\documentstyle[floats,epsf,aps]{revtex}

\begin{document}
\bigskip
\newcommand{\beq}{\begin{equation}}
\newcommand{\eeq}{\end{equation}}
\large
\title{
{\bf Dynamical fragment production in central collisions Xe(50 A.MeV)+Sn} } 
\author{ {\bf Regina Nebauer$^{*+}$ and J\"org Aichelin$^*$} 
\\
{\normalsize
$^*$SUBATECH}\\
{\normalsize
 Universit\'e de Nantes, EMN, IN2P3/CNRS}\\
{\normalsize    4, Rue Alfred Kastler, 44070 Nantes Cedex 03, France }\\
{\normalsize $^*$ Universit\"at Rostock, Germany}}

\maketitle

\

{\Large
\begin{abstract} \large
{\bf Abstract:} For central collisions Xe(50 A.MeV)+Sn we compared experimental data from the INDRA
detector with QMD simulations. Theory as well as experiment show a clear binary
character of the fragment emission even for very central collisions. From the time
evolution of the reaction (QMD simulation) we could built up a scenario for the 
dynamical emission of fragments
\end{abstract}

\section{introduction}

It is known since long that for almost all particles observed in heavy ion
reactions between 30~A.MeV and 200~A.GeV  
the transverse kinetic energy spectra have a
Maxwell-Boltzmann form, predicted for an emission from an equilibrated source.
However, the apparent temperature of the spectra and hence the average 
kinetic energy of the particles is quite different for different hadrons and
fragments and increases with increasing mass and increasing energy. This
observation seemed to exclude an identification of the apparent temperature 
with a real temperature of the system.

Recently it has been conjectured \cite{pbm,mar,reis} that at all energies 
between 50 A.MeV and 200~A.GeV the assumption of a strong radial flow can 
reconcile the mass dependence of the apparent temperature with 
thermodynamics.  At relativistic
and ultra-relativistic energies this has been inferred by comparing transverse
pion, kaon and proton spectra \cite{pbm}. At energies below
500~A.MeV the lever arm is still larger because one can include the 
intermediate
mass fragments (IMF's) of masses in between 2 and 10 \cite{mar,reis}, emitted
at midrapidity, to separate radial flow and temperature. The deviations in forward 
and backward direction are usually interpreted as preequilibrium emission.
We will show that at low energies
the increase of the transverse kinetic energy as a function of the mass 
of the fragment is caused
by a mechanism already proposed many years ago by Goldhaber \cite{gol}. 
He showed that if multifragmentation is a sudden break off of the
fragments 
the nucleons retain their momentum due to the Fermi motion and one expects a
variance of the momentum distribution of the fragments which increases 
linear proportional to A. 
The process proposed by Goldhaber is exactly the opposite of thermal equilibration.
There multifragmentation occurs after the system has reached global
equilibrium and is a process which is sufficiently slow 
to retain that global equilibrium until the moment of break off. 
Experimentally both processes are very difficult to disentangle, the simulation
programs, however, which reproduce the final kinetic energy distribution,
allow to address this question. In the process proposed by Goldhaber the
kinetic energy of the final fragments is already initially present as 
kinetic energy of the
nucleons which finally form the fragment. On the contrary, in thermodynamical
processes the temperature of the system and hence the kinetic energy is created during 
the interaction between the heavy ions.

For our study we use simulations for the reaction 
Xe(50~A.MeV) + Sn
which has recently been studied using 
the INDRA detector at GANIL. This detector has been constructed to study
multifragmentation and therefore the angular coverage and the 
energy thresholds
have been chosen to be better than that of any other $4\pi$ detector 
elsewhere. Hence the data taken with this detector are most suitable 
to confirm or disprove the theories embedded in the simulation programs. 
A detailed comparison of our results with the experimental data will
be published elsewhere. Here we mention that not only the mass dependence
of the average kinetic energy but also the kinetic energy spectra
are in reasonable agreement with experiment.

For details about the QMD approach we refer to reference \cite{aic}. In
this program the nucleons are represented by Gaussian wave packets with a
constant width. The time evolution of the centers of these wave packets
is given by Euler Lagrange equations derived from the Lagrangian of the system.
   

\section{Experimental results}

We selected central collisions from the INDRA data Xe(50 A.MeV)+Sn with the conditions
of completeness and high transverse energy of the light particles. Completeness means
that 80\% of the total charge of the system and 80\% of the initial longitudinal 
momentum are detected \cite{mar}. The centrality of the collision is given by the total 
transverse
energy of light particles ($Z\le 2$) \cite{luka}. We selected collisions with $E_{trans}\ge
450 MeV$, in the QMD simulation this corresponds to a reduced impact parameter
$b/b_{max} = 0.3$ 

We focus on the production of {\bf intermediate mass fragments ($Z\ge 3$)} and investigate
the angular dependence of the fragment emission.
In the center of mass system the experiment shows a flat
angular distribution ($dN/dcos\theta_{cm}$) between
$60^\circ \leq \theta_{cm} \leq 120^\circ$ as well as 
a constant average kinetic 
energy  for fragments $Z\geq3$. 
In forward and backward direction a strongly enhanced cross section 
is observed. The INDRA collaboration made use of this observation and
presented their data in two angular bins:
$60^o~\le~\theta_{CM}~\le~120^o$ (IMF's emitted in this angular range are called
mid-rapidity fragments (MRF's)) and $\theta_{CM}~<~60^o,~\theta_{CM}~>~120^o$ called
projectile/target like fragments (PTF's).

If we take into account the above mentioned differences and plot the average
kinetic energy separately for the fragments emitted in forward/backward
direction and in the central region, we find two different slopes.
In the forward/backward direction we have a linear rise. The kinetic
energies of fragments emitted in the central region is much lower. The linear
rise is much less pronounced and vanishes for fragments with a charge greater than
12. Fragments emitted from a purely thermal source show, independent of the mass, i.e.
the charge, a constant energy $\langle E_{kin}\rangle=\frac{3}{2}T$. Obviously, this is
not the case. May be that the temperature increases with the mass of the fragment or a
collective flow of the nucleons is present. Adding to a thermal system a flow component
per nucleon, we obtain a
linear rise of the kinetic energy with increasing mass. The aim of our work is to find
an explanation for this linear increase of the kinetic energy with the fragment mass,
first of all in the midrapidity zone.
   
\begin{figure}[h]
\vspace{-2.5cm}
\epsfxsize=15.cm
\epsfbox{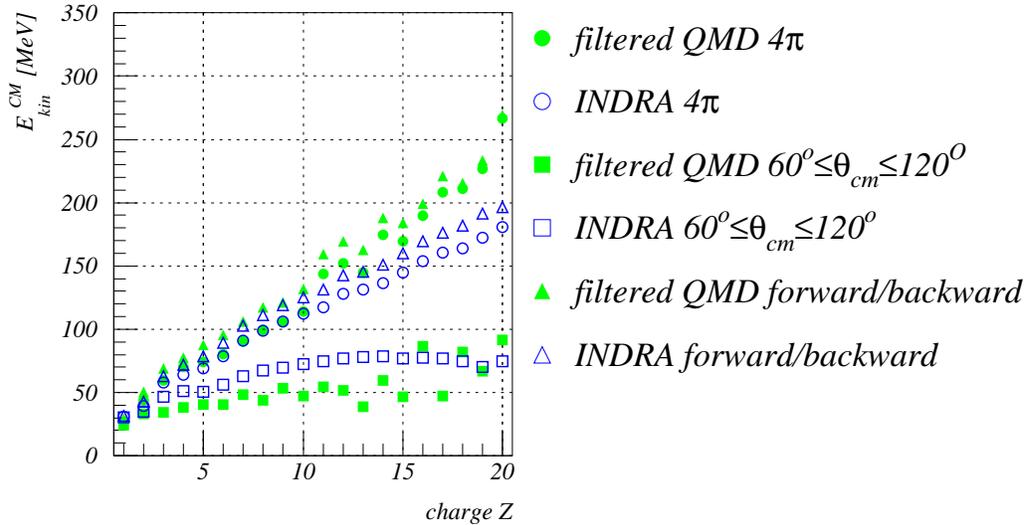}
\vspace{-1cm}
\caption{\textit{Average kinetic energy, the error bars are suppressed}}
\label{ekinz}
\end{figure}

To find out whether the linear increase is due to a flow component, we have to analyze
the energy spectra. 
First we compare the energy spectra for 
fragments emitted in the forward/backward zone for filtered QMD and INDRA data, 
figure~\ref{spfb}. We displayed three typical spectra for different charges, Z=5, 10 and
18. On the left hand side we show the spectra, on the right hand side 
the difference between the experimental and QMD spectra.
With increasing mass (charge) the maximum
of the spectra is shifted towards higher energies.
An  explanation for this shift 
comes from the strong binary character of the reaction.
We indicated the beam energy and one can see that the spectra display even
for this central collisions the beam energy, the nuclei show a certain degree of
transparency.

\begin{figure}[h]
\vspace{-1cm}
\epsfxsize=15.cm
$$
\epsfbox{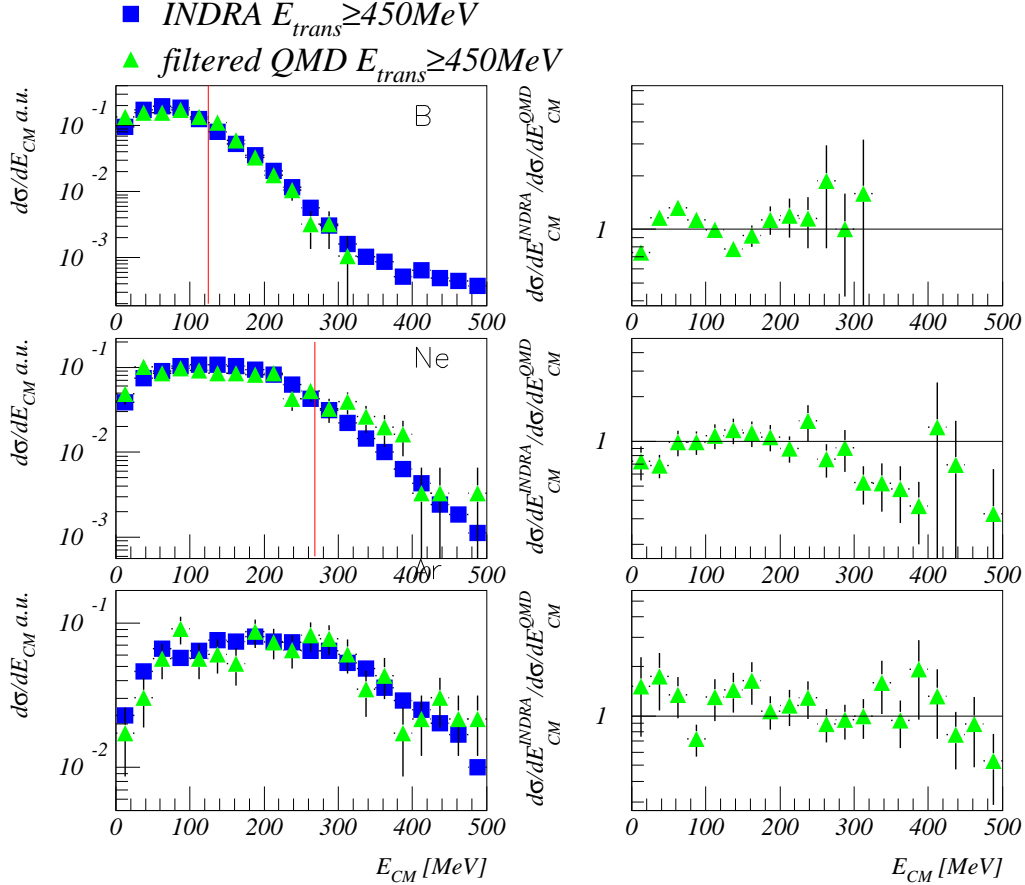}
$$
\vspace{-2cm}
\caption{\textit{Energy spectra for INDRA  and QMD data in forward/backward
direction in the center of mass frame. On the right side we display the logarithm
of the ratio of the two spectra.}}
\label{spfb}
\end{figure}

This general shape of the spectra is quite well described by QMD.
Even for big fragments the low energy domain of the spectra corresponds to the
experimental data. A reassuring fact for the hypothesis of the transparency of the 
nuclei if we assume that the suppression of low energies comes from the Coulomb
repulsion. 
Nevertheless the high energy domain is overestimated by QMD for very big fragments, the
average kinetic energy
is higher in the QMD calculation than for INDRA data for fragments $Z>10$. This
difference increases with $Z$. That we have fragments with a
lower kinetic energy in the INDRA data than in the QMD calculations shows
that the nuclei are less ``transparent'' in reality than predicted by the model.

As the emission in midrapidity is weak for
the QMD model and vanishes nearly for charges higher than twenty, we do not have
enough statistics at our disposal for big fragments. A comparison of the spectra
 is only
possible for charges up to twelve, for higher charges the fluctuations 
render the analysis meaningless, for the average kinetic energy as
well as for the spectra. Focusing on the INDRA spectra, we constate no flow component.
In the case of a radial flow we would observe a maximum in the spectra which is shifted
to higher energies with increasing mass. In contrary, we constate a change of the slope
of the spectra. Thus. the increase of the mean kinetic energy with the fragment mass is
not due to a radial flow but to a change of the slope.

\begin{figure}[h]
\epsfxsize=15.cm
$$
\epsfbox{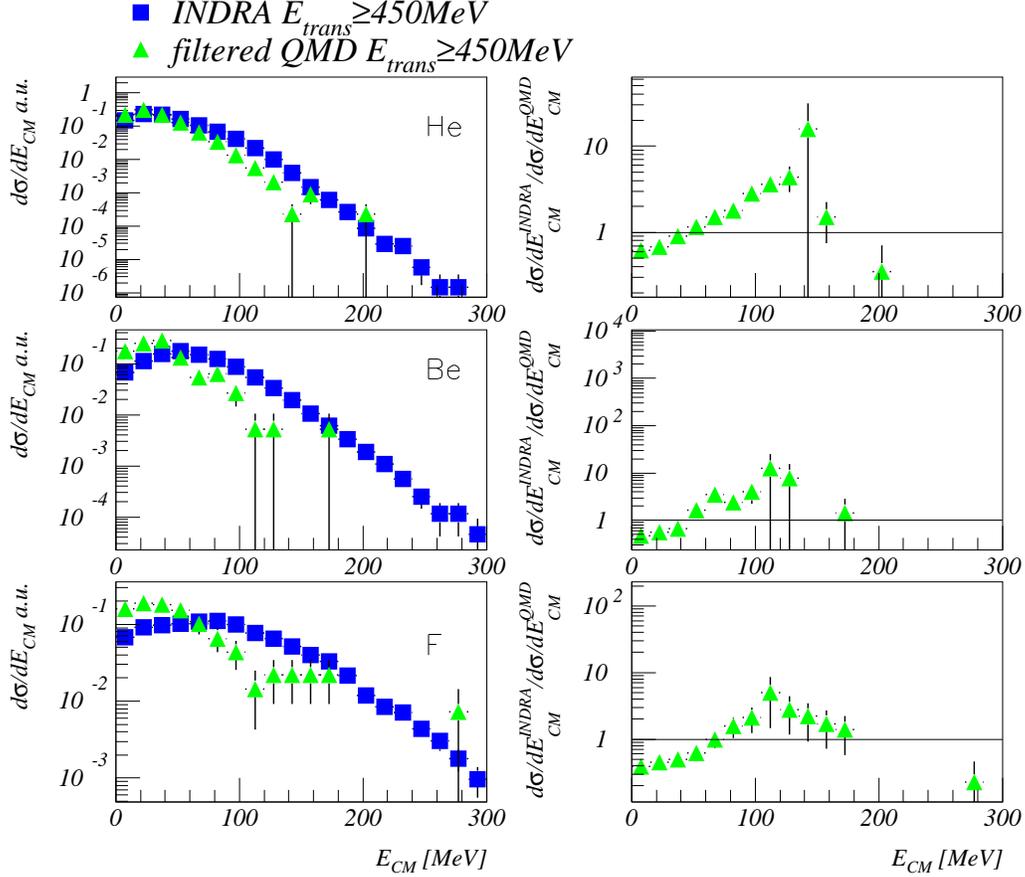}
$$
\vspace{-2cm}
\caption{\textit{Energy spectra: we compare the QMD and INDRA data. On the right 
side the
spectra are displayed, on the left side the surprisal analysis.}}
\label{spsouq}
\end{figure}

For light particles the agreement between the INDRA data and theory is quite
good, for charges $Z>3$ the low energy part is overestimated by QMD. Due to this low
energies most of the fragments are lost in the filter. Knowing the origin of the
kinetic energies we would find an explanation why it is underestimated by QMD.

\section{Time evolution of the reaction}

From the comparison in the previous section we found a reasonable agreement between
the model calculations and experimental data. In order to find out the origin of the
above described fragment properties we investigate the time evolution of the reaction
from the QMD model.

As the QMD model works with effective charges, we redefine the intermediate mass
fragments with the mass:
\begin{itemize}
\item {\bf{intermediate mass fragments (IMF):}} mass of the fragments $A\geq5$
\end{itemize}

A first general idea of the time evolution of the collision can be obtained 
from the time
evolution of the density of the system.
If the maximal density is reached, 
the nuclei have their maximal overlap, after that the system
expands and the density decreases. This permits to find the time scale of 
the reaction. 

The total density is 
the sum over all nucleons which are described by Gaussians:
\begin{equation}
\label{densdef}
\rho(\vec{r},t)
\propto\sum_{i=1}^{A}
e^{-\frac{(\vec{r}-\vec{r_i}(t))^2}{2{L}}}
\label{denga}
\end{equation}
The width of the Gaussians is $4L=4.33~fm^2$ and A is the number of nucleons
present in the system.

In figure~\ref{dic}, left hand side, we plot the time evolution of the 
total density in the center of the reaction, $\vec{r}=0~fm$. 
The maximum density is obtained at $\approx 50~fm/c$, on 
the same time scale the
system expands and reaches at $120~fm/c$ a low density phase where the 
fragments do not interact anymore. 
\begin{figure}[h]
\vspace{-3.cm}
\epsfxsize=17.cm
$$
\epsfbox{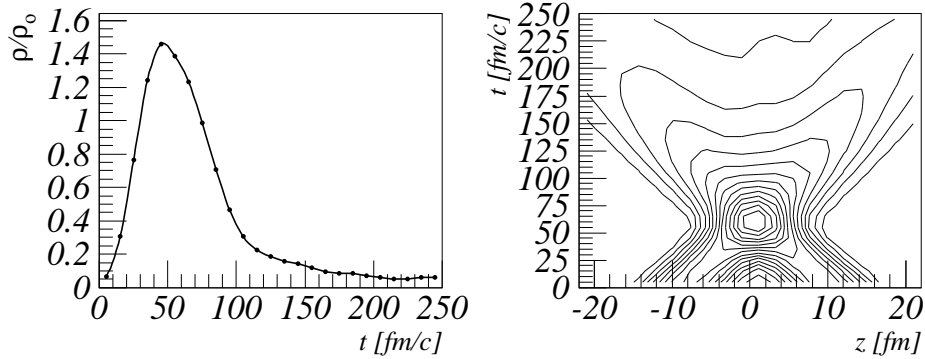}
$$
\vspace{-3cm}
\caption{\textit{Time evolution of the density (left) and of the density profile along
the beam (z) axis (right) for the system Xe(50~A.MeV) + Sn ,b=3~fm.}}
\label{dic}
\end{figure}

On the right hand side of the same figure we display the density profile along the
beam (z) axis. Here we can follow the two nuclei, they occupy the same coordinate 
space at
$50..60~fm/c$. The system expands after $120~fm/c$. We find  
that this quasi-central ($b= 3~fm$) collision is semi transparent. Projectile
and target pass each other without being seriously decelerated.
For $b=0~fm$ we get the same result.  
That binary character is confirmed by experiment.

In our theoretical analysis we follow the separation presented in the previous section.
We start out with the time
evolution of the average longitudinal and transverse momentum of the nucleons entrained
finally in 
intermediate mass fragments (IMF's) $A\geq 5$. Both are displayed in 
figure~\ref{ptrpzt}. The upper graphs show for three different impact parameters
the average longitudinal and 
transverse momentum of the MRF's, 
the lower graphs that of the PTF's.

\begin{figure}[h]
\vspace{-2.5cm}
\epsfxsize=17.cm
$$
\epsfbox{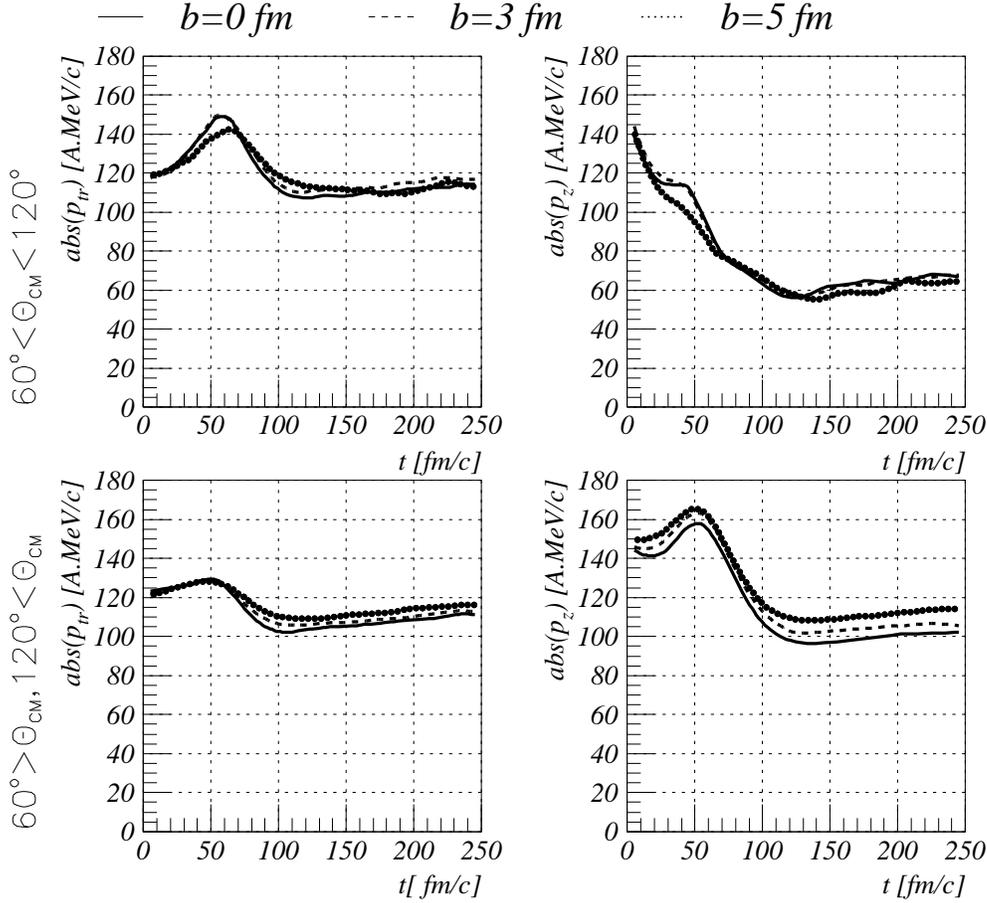}
$$
\vspace{-1cm}
\caption{\textit{Time evolution of the transverse and longitudinal momentum for
nucleons
emitted in IMF's at
midrapidity (first row) and in forward/backward (second row)}}
\label{ptrpzt}
\end{figure}

First of all we observe that all classes of fragments have the same 
average initial transverse
momentum. Consequently the initial - final state correlations in momentum 
space are small. This initial value is a consequence of the Fermi motion 
of the nucleons. 
 
It is the seminal result of this work that the average transverse momentum does not
change considerably during the reaction as one would expect 
if the system equilibrates and heats up by converting beam energy into thermal
heat.
Thus the final energy distribution cannot be associated with a temperature but is
merely
a reflection of the initial Fermi distribution of the nucleons. 
This explains also
why the apparent temperature of the fragment kinetic energy spectra 
($\approx 15$ MeV) is large as compared
to the temperature extract from the caloric curve ( $\approx 5$ MeV).
We observe as well that the impact parameter dependence is weak.

If one assumes that multifragmentation is a fast process where the nucleons,
entrained in a fragment, separate from the rest of the system that fast
that they retain their initial momentum one can calculate
\cite{gol} the expected average kinetic energy of the fragments. It is
equivalent to that one obtains if one picks randomly N nucleons 
out of A which have a Fermi distribution with $\sum_i^A p_i = 0$. 
One observes that the average momentum of the N
nucleons and hence the mean momentum of the fragment is zero but also 
a variance of the momentum distribution of 
$$<(\sum_i^N p_i)^2> = {3k_{Fermi}^2 \over 5} N {A-N \over A -1}$$ 
where A is the size of the system. Consequently, for this process a linear
dependence of the mean fragment kinetic energy on the fragment mass is
expected for small values of N.

The nucleons finally emitted as MRF's loose their
longitudinal momenta in three steps. Very early in the reaction
collisions reduce the relative longitudinal momentum between the nucleons 
from projectile and target while the high density zone is created. The nucleons
move now towards this high density zone and loose longitudinal momentum
while climbing up the potential wall. Finally the fragments separate which
decreases a third time the longitudinal momentum. 

The group of nucleons emitted finally as PTF's
passes the reaction zone without being really affected.
Note that at this small impact parameter there are almost no spectators.
 Hence, at $50~A.MeV$ the 
nuclei are semi-transparent. How this is possible we discuss later.
At lower energies ($<10~A.MeV$) we find for small impact 
parameters 
the formation of a compound nucleus and hence an equilibration of projectile and 
target nucleons and at higher energies the formation of a mid-rapidity fireball. Thus
at lower
and higher energies the stopping is more complete than at that intermediate 
energy. We analyzed in the same way the average momenta of the prefragments and observe
that
the final momentum of the fragments is almost identical with the initial momentum of
the group of nucleons which finally will form a fragment.

The nucleons interact via the potential
\beq
V=-70\frac{\rho}{\rho_0} +120(\frac{\rho}{\rho_0})^2
\eeq
where the density is given by equation~\ref{denga} and $\rho_0$ is the normal
nuclear matter density.
In order to reveal the physics which drives the reaction we display the relative
density of
those nucleons which are finally entrained in MRF's or PTF's as a function of time in
the x-z plane
\beq
\rho^{\tiny MRF/PTF}_{rel}(x,z,t)={\rho_{\tiny MRF/PTF}(x,z,t) \over
\rho_{total}(x,z,t)}
\eeq 
and superimpose the gradient of the potential in the x-z plane as arrows where
x is the direction of the impact parameter. 
For the sake of a clearer display we plot nucleons coming
from the projectile only.

The motion of the nucleons in the potential of a nucleus is a sequence of acceleration
and deceleration. Nucleons on the surface are almost at rest, due to the density (and
thus the potential) gradient they become accelerated towards the center of the nucleus.
They
reach their maximal momentum when they pass
the center of the nucleus, climb up the potential on the 
other side and are finally
at rest again when arriving at the surface. 
When a heavy ion collision occurs, the position of
the nucleons in the projectile or target determines whether they "feel"
the heavy ion collision right from the beginning or only when the
high density phase has already passed. We will show that the initial position 
of the entrained nucleons decides 
as
well whether the fragment is finally observed at midrapidity or in forward/
backward direction.

\begin{figure}[h]
$$
\epsfxsize=15.cm
\epsfbox{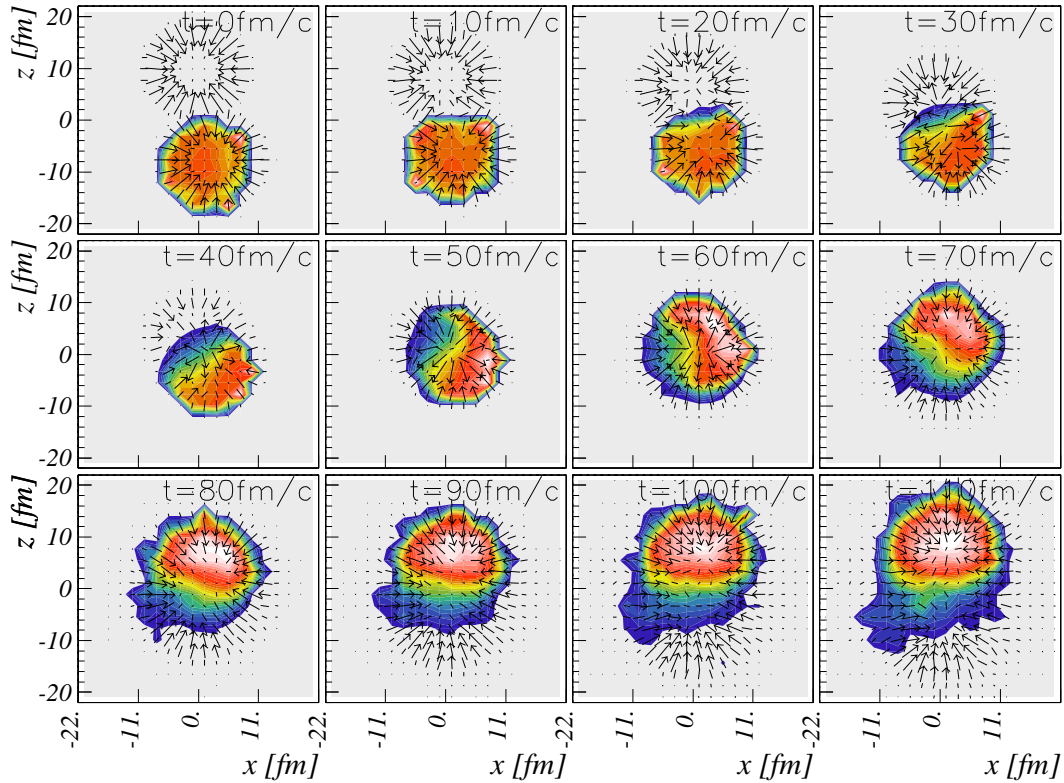}
$$
\vspace{-.5cm}
\caption{\textit{Movement of the nucleons finally emitted as IMF's in forward/backward
direction in the mean field potential for collisions at impact parameter $b=3~fm$. We 
display the fraction of these nucleons on the
total density (shadow) and the gradient of the potential (arrows) projected on the x-z
plane}}
\label{potgfg3}
\end{figure}

In figure~\ref{potgfg3} we display the motion of the nucleons finally entrained in
PTF's for a reaction at  $b=3~fm$. The spatial distribution of those nucleons 
is almost identical with
that of all nucleons present in the projectile.
In the first step of the collision the nucleons move away from the target
into the yet unperturbed part of the projectile. When they arrive
at the back end of the projectile they invert the direction of their momenta. 
They are then accelerated in longitudinal direction towards the 
center of the reaction.
As we have a finite impact parameter, asymmetry 
effects occur.  When the high density phase occurs, the nucleons take the line
of
least resistance, i.e. they follow the minimum of the potential on the right hand
side (for the projectile nucleons, the target nucleons take the inverse direction on
the
other side). The larger part of the nucleons pass the reaction center when the
potential barrier has disappeared. At zero impact parameter this passage of the
reaction center at lower densities is more clear. 
Hence the nucleons pass the center without a larger change of their initial 
momentum. The initial correlations \cite{goss} among the nucleons which
finally form a fragment survive the reaction because
all potential gradients are small.

\begin{figure}[h]
$$
\epsfxsize=15.cm
\epsfbox{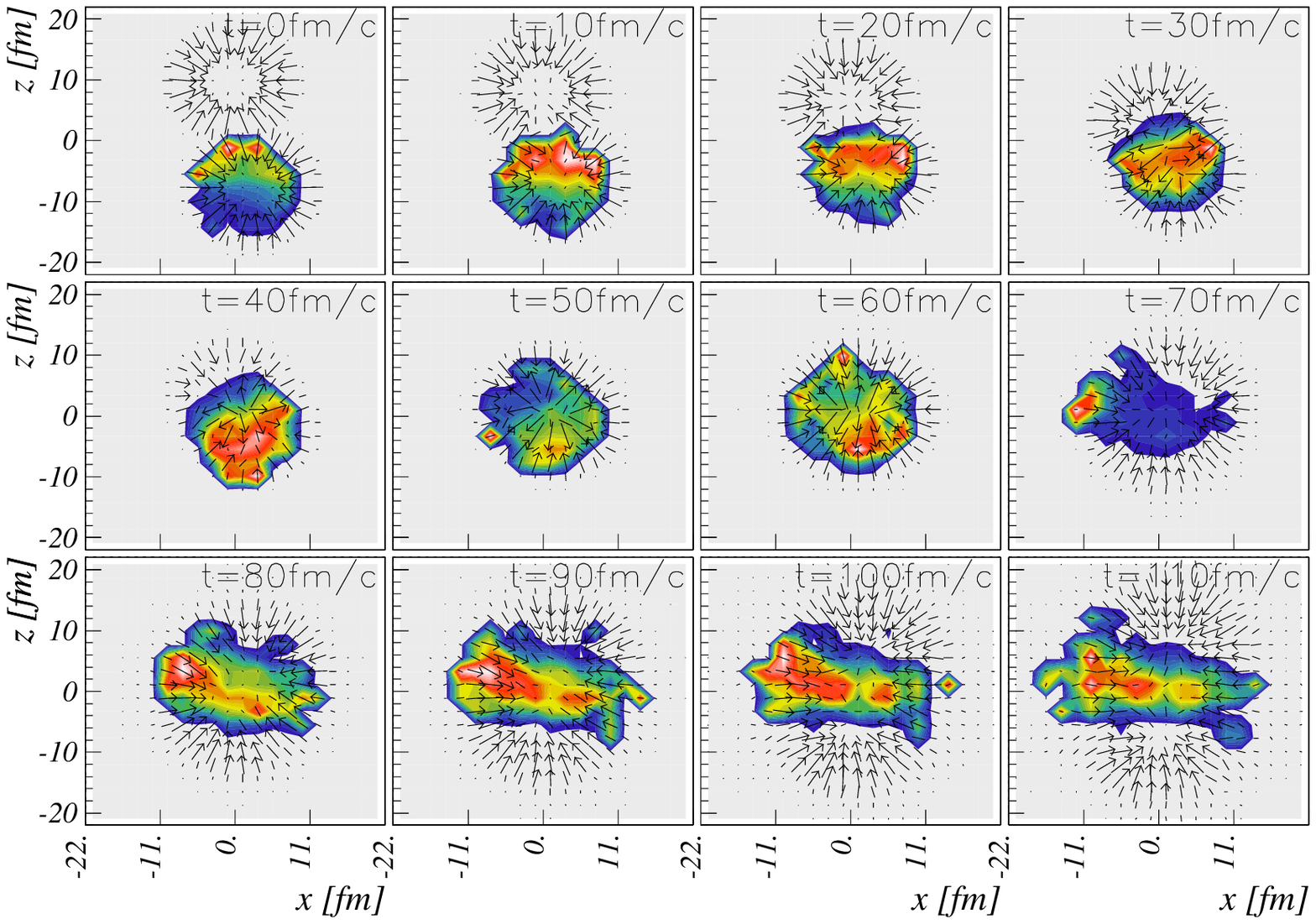}
$$
\vspace{-.5cm}
\caption{\textit{Movement of the nucleons finally emitted as IMF's in midrapidity in
the
mean field potential for collisions at impact parameter $b=3~fm$. We display the
fraction 
of these nucleons on the
total density (shadow) and the gradient of the potential (arrows) projected on the x-z
plane}}
\label{potgm3}
\end{figure}

Nucleons finally emitted as MRF's (fig. \ref{potgm3}) are strongly
located at the front end of the nuclei. These are the nucleons which 
climbed up the 
nuclear potential before they are at rest on the top of the potential wall. 
Due to their position they are involved in 
the collisions between projectile and target nucleons right from the beginning. 
This supports the
deceleration. (Later collisions are to a large extend Pauli suppressed.) 
They escape the barrier 
in transverse direction. 
As their momentum (longitudinal as well as 
transverse)
is quite small, the nucleons stay longer in the
center of the reaction which favors the mixing of projectile and target 
nucleons. 
When leaving the reaction zone the fragments become decelerated due to the
potential interaction with the rest of the system. This 
deceleration balances
the gain in energy due to the prior acceleration in transverse direction, 
although the
physics of both processes is rather independent.

\section{Conclusion}

In conclusion, the multifragmentation process in central collisions Xe(50 A.MeV)+Sn has
been studied. Experimental data from the INDRA detector at GANIL were compared with 
QMD simulations. From the angular dependence of the emission we could define two
regions where the reaction mechanism are obviously different. 
A clear projectile/target
like character could be observed in forward/backward direction even for central
collisions.  
From the time evolution we have found that in agreement with
experiment even in central collisions the reaction is semi transparent. 

For the emission in midrapidity the linear increase of the fragment 
kinetic energy with the fragment mass for small fragment masses
finds its natural explication in terms of the initial Fermi motion.
When the fragments separate fast from the system this linear dependence is
expected. This as well as the binary character of the reaction points towards
the conjecture that a heavy ion collision at this energy is a fast process
and does not pass a state of global equilibrium.

\end{document}